\documentclass[twocolumn,showpacs,preprintnumbers,amsmath,amssymb,superscriptaddress]{revtex4-1}
\usepackage{amsmath,amssymb,amsfonts}
\usepackage{graphics,subfigure}
\usepackage{epsfig}
\usepackage{color}
\usepackage{graphicx}
\usepackage{dcolumn}
\usepackage{bm,hyperref}

\renewcommand{\eqref}[1]{Eq.~(\ref{#1})}

\begin{document}
   \title{Criterion for explosive percolation transitions on complex networks}
\author{Hans Hooyberghs}
\email{hans.hooyberghs@fys.kuleuven.be}
\affiliation{Instituut voor Theoretische Fysica, Katholieke 
Universiteit Leuven, Celestijnenlaan 200D,  B-3001 Leuven, Belgium}
\author{Bert Van Schaeybroeck}
\affiliation{ Koninklijk Metereologisch Instituut (KMI), Ringlaan 3,
B-1180 Brussels, Belgium}
\begin{abstract}
 In a recent Letter, Friedman and Landsberg  discussed the underlying mechanism of explosive phase transitions on complex networks [Phys. Rev. Lett. \textbf{103}, 255701 (2009)]. This Brief Report presents a modest, though more insightful extension of their arguments. We discuss the implications of their results on the cluster-size distribution and deduce that, under general conditions, the percolation transition will be explosive if the mean number of nodes per cluster diverges in the thermodynamic limit and prior to the transition threshold. In other words, if, upon increase of the network size $n$ the amount of clusters in the network does not grow proportionally to $n$, the percolation transition is explosive. Simulations and analytical calculations on various models support our findings.
\end{abstract}

\maketitle
The percolation problem, which deals with the structure and connectivity of a network or lattice as some of its nodes or edges are removed,  is one of the most widely studied problems in network theory \cite{stauffer,dorogovtsev}.  During the last decade, much attention was devoted to the percolation transition on complex network, which has amongst others applications pertaining to virus spreading  \cite{kenah2007} and network failures \cite{cohen,percol}.
 
Recently, Achlioptas et al.~have discovered a new class of percolation transitions~\cite{achiel}. The transition which they term \textit{explosive} is marked by a vanishing ``width'' in the thermodynamic limit.
According to the original Achlioptas process, two candidate edges are selected at random at each timestep. The edge resulting in the smallest product of its connecting cluster sizes is then effectively laid. Following the work of Achlioptas et al., explosive percolations have been observed on various networks and lattices using a wide variety of edge-addition rules, all of which involve the knowledge of the cluster sizes~\cite{friedman2009,ziff2009,radicchi2009,souza2010,cho2009,manna2009,moreira}. 

An important step towards a more profound understanding of explosive transitions was made by Friedman and Landsberg~\cite{friedman2009}. They put forward the criterion that a necessary condition for the appearance of an explosive transition is the existence of a nonzero fraction of nodes in the powder keg in the thermodynamic limit. In a network of size $n$, the powder keg $F(t(n^\sigma),n^{1-\beta})$ quantifies the number of nodes in clusters of size larger than $n^{1-\beta}$ at time $t(n^\sigma)$ when the first cluster of size $n^\sigma$ makes its appearance. A trivial constraint is $0<1-\beta<\sigma<1$. In this Brief Report, we attempt to gain additional insight by presenting a more simplified criterion. Therefore, we consider the implications of a nonzero fraction of nodes in the powder keg on the cluster-size distribution for explosive transitions.

We assume that, close to the phase transition in the non-percolated regime, the number of clusters of size $s$, here denoted $n_s$, is approximately described by a power law with a cut-off size $s^*$ for large $s$. Thus 
\begin{equation}\label{ansatz}
 n_s = A s^{-\tau}e^{-s/s^*}\hspace{0.1cm} \textrm{ for  } s\rightarrow \infty.
\end{equation}
Here the normalization constant $A$ is fixed by the constraint $n = \sum_s sn_s$ and, as occurs also for second-order percolation transitions~\cite{stauffer}, the cut-off length scales as a power of the network size; in other words $s^*\propto n^\varsigma$ with $\varsigma$ a non-negative exponent. Cluster-size distributions of the form of \eqref{ansatz} are for instance encountered for explosive transitions in simulations on the Achlioptas process on scale-free networks \cite{radicchi2009}, on the model of D'Souza et al.~\cite{souza2010} and on the model of Cho et al.~\cite{cho2009,manna2009}, as well as in the theoretical work of Da Costa et al.~\cite{costa2010}. However, some studies reveal deviating cluster-size distributions~\cite{pan,chen,araujo,nagler}; we will discuss them later on. 

Converting sums over the cluster-size distribution into integrals, the powder keg is given by
\begin{equation}\label{Feq}
 F(t(n^\sigma),n^{1-\beta}) \approx \int^{n}_{n^{1-\beta}} sn_s \:ds .
\end{equation}
For large networks, the integral can be evaluated analytically using the preceding assumptions and asymptotic expansions of the (incomplete) gamma function.
We obtain:
\begin{equation}
\frac{1}{n}F(t(n^\sigma),n^{1-\beta}) \propto n^{(1-\beta)(2-\tau)}\quad \textrm{if}\quad \tau>2.
 \end{equation}
The fraction of nodes in the powder keg thus vanishes in the thermodynamic limit if $\tau>2$. Conversely:
\begin{equation}
F(t(n^\sigma),n^{1-\beta}) \propto n\quad \textrm{if}\quad \tau\leq2,
\end{equation}
in other words, the powder keg contains a nonzero fraction of nodes. In both regimes, the results are verified numerically by an explicit evaluation of the integral in \eqref{Feq}. Note that the conclusions are independent of the value of  $\varsigma$ and the exact definition of the upper and lower boundary of the  powder keg, quantified by the exponents $\sigma$ and $\beta$. Therefore, we establish the criterion that the percolation transition is of the explosive type if  $\tau \leq 2$ at the onset of the phase transition. 

The actual determination of $\tau$ in simulations is rather difficult. As an alternative, we present an equivalent criterion in terms of the mean number of nodes per cluster at the onset of the phase transition, $\langle s \rangle_o $, the latter being a more easily accessible quantity in simulations. Let $n_c$ be the number of clusters, then $\langle s \rangle_o = n/n_c$. Since $n_c = \sum_s n_s$, the mean cluster size $\langle s \rangle_o$ can be evaluated analytically using~\eqref{ansatz}. We deduce that in the thermodynamic limit $\langle s \rangle_o$ diverges if and only if $\tau \leq 2$, a statement which can again be verified numerically. In other words, a percolation transition will be explosive if the mean number of nodes per cluster diverges \emph{at the onset} of the phase transition, i.e.~\emph{before} the giant cluster is formed. 
For non-explosive transitions, a divergence of $\langle s \rangle_o$ can solely be caused by the formation of a giant cluster.

Our criterion involving $\langle s \rangle_o$ quantifies the result of the intuitive mechanism underlying explosive processes known thus far: they all aim at avoiding any cluster size to become much larger than any other, therefore giving rise to a network with a small amount of clusters, all of which have more or less the same size~\cite{cho2009,manna2009,moreira}. Quantitatively, the transition will be explosive if the cluster density $n_c/n$ vanishes in the thermodynamic limit, i.e.~if, upon increase of the network size $n$, the number of clusters does not grow proportional to $n$.
\newline\newline\begin{figure}[t!]
\includegraphics[height = 0.5\textwidth,angle=270]{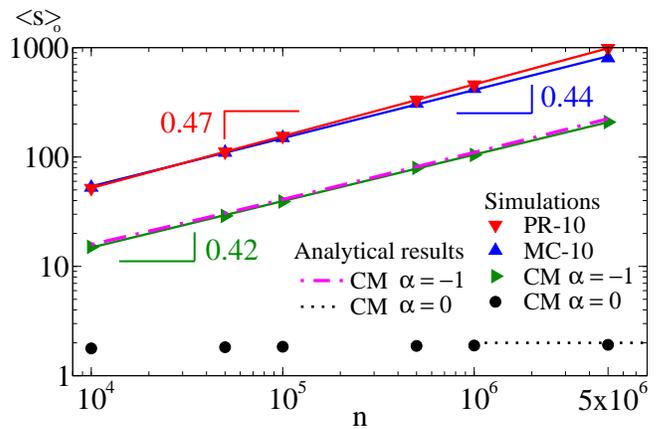}
\caption{(Color online) Log-log plot of $\langle s \rangle_o$, the mean number of nodes per cluster at the onset of the phase  transition, as a function of the network size for various percolation processes. The onset of the phase transition  coincides with the time when the first cluster of size $\sqrt{n}$ appears. For the Cho-Manna process (CM), results for an explosive regime ($\alpha = -1$, green right  triangles) and a non-explosive regime ($\alpha = 0$, black circles) are shown. The (black) dotted line shows the analytically obtained asymptotic limit for the non-explosive regime, while the (magenta) dot-dashed line gives the analytical result for the explosive regime. Also shown are simulation results for the explosive min-cluster sum rule (MC-10, blue up triangles) and the explosive extended-product rule (PR-10, red down triangles).
All data points are averages over 20 realizations of the percolation process.\label{figuur}}
\end{figure}
We have checked our criterion, both numerically and analytically in various explosive models. Firstly, we examined the percolation process suggested  
by Cho et al.~\cite{cho2009} and Manna et al.~\cite{manna2009}, in which the cluster-size distribution follows approximately a power-law with a cut-off according to \eqref{ansatz}. The starting point is a network consisting of $n$ nodes without links.
At each timestep, two clusters are selected, each with probability proportional to $s^{\alpha+1}$ with $s$ their cluster size and $\alpha \in [-1,0]$. The two selected clusters may be the same. A link is then laid between two random nodes of the selected clusters. Self-linking and multiple links between the same nodes are prohibited. The ensuing percolation transition was found to be explosive if $\alpha <-0.5$ and non-explosive otherwise \cite{cho2009,manna2009}. Fig.~\ref{figuur} shows the mean number of nodes in a cluster at the appearance of  the first cluster of size $\sqrt{n}$, which is generally considered as the onset of the phase transition. 

The non-explosive regime with $\alpha = 0$ corresponds to random percolation.  An expression for the cluster-size distribution for random percolation was obtained in Ref.~\cite{ziffoud}. An exact computation of the mean cluster size per cluster in the thermodynamic limit then reveals the finite asymptotic limit $\langle s\rangle_o = 2$, which nicely corresponds with our simulation data  as seen by the dots and dotted line in Fig.~\ref{figuur}.

Also in the explosive regime, $\alpha = -1$, the cluster-size distribution can be obtained analytically \cite{cho2009}. At time $t$, 
\begin{equation}
 n_s(t) = n(1-t)^2t^{s-1},
\end{equation}
and thus 
\begin{equation}
\langle s \rangle =\frac{1}{1-t}\label{smol}.
\end{equation} 
An expression for the mean number of nodes per cluster \emph{at the onset} of the phase transition (i.e. when there is exactly one cluster size larger than $n^{1-\beta}$), is then found by solving the equation
\begin{equation}
1 = \sum_{s=n^{1-\beta}}^\infty n_s(t)
\end{equation}
for $t$ and introducing the result in \eqref{smol}. Results of a numerical evaluation of this procedure for $\beta = 0.5$ are shown in Fig.~\ref{figuur}, which also shows the value of  $\langle s\rangle_o$  determined in the simulations. The theory is found to agree well with the simulations. Both reveal a mean cluster size per cluster which diverges upon approach of the thermodynamic limit as  $\langle s \rangle_o \propto n^\gamma$ with $\gamma = 0.42 \pm 0.02$. We conclude that simulations and analytical computations on the Cho-Manna model confirm our second criterion.
\newline\newline
Our presented analysis and simulations so far relied on the ansatz for the cluster-size distribution, \eqref{ansatz}.
Based on the cluster-size distribution near criticality, different types of models exhibiting explosive percolation may be distinguished. A power-law distribution with a bump is observed in the models of Refs.~\cite{pan,chen}, while a lattice model  yields a Gaussian cluster-size distribution \cite{araujo}. On the other hand, there exists an extremely explosive model in which the cluster size distribution is not continuous but highly discrete. In the latter model, at  each timestep maximally two cluster-sizes are present in the network~\cite{nagler}.

We have performed simulations on two models in which the cluster-size distribution features a bump for large cluster sizes. Both models start from $n$ distinct nodes. At each timestip, ten possible edges are selected at random. In the first model, a straightforward extension of the rule of Achlioptas denoted as PR-10, the edge which results in the smallest product of its connecting cluster sizes is effectively laid. In the min-cluster sum rule (MC-10), out of ten edges, the edge which minimizes the size of the component formed if the edge is occupied, is laid. The cluster-size distribution in the non-percolating regime does not follow exactly \eqref{ansatz} \cite{pan}: although it is characterized by a power-law behavior for small cluster sizes and an exponential cut-off for large sizes, there is a pronounced bump in the probability distribution for intermediate sizes. Nevertheless, as shown in Fig.~\ref{figuur}, for both processes the mean number of nodes per cluster follows a diverging power-law of the network size,  $\langle s \rangle_o \propto n^\gamma$. For the MC-10 rule, we obtain $\gamma = 0.44\pm0.02$; while $\gamma = 0.47\pm 0.01$ for the extended product rule. Although the derivation presented in this Report is based on \eqref{ansatz}, it seems that our criterion holds in explosive models in which the ansatz is no longer true. In our opinion, the models with a bumped cluster-size distribution still satisfy our criterion since their distributions contain two essential ingredients of distribution \eqref{ansatz}, that is, a power-law distribution at low cluster-size and an exponential cut-off length which diverges in the thermodynamic limit.
 
We have also tested our criterion using a model with a highly discrete cluster-size distribution, based on an extreme form of the Achlioptas process. The PR-$\infty$ rule takes into account all possible links and then selects the link which results in the smallest product of its connecting cluster sizes. Suppose for simplicity the system size $n$ to be an integer power of two. The cluster-size distribution is then trivial since maximally two types of clusters exist at each timestep \cite{nagler}. Suppose that $2^\xi$ is the largest power of two smaller than $n^{1-\beta}$, thus $2^\xi < n^{1-\beta} < 2^{\xi+1}$. At the moment of the first formation of a cluster of size $n^{1-\beta}$ or larger, there is a single cluster of size $2^{\xi +1}$, while all other nodes belong to clusters of size $2^\xi$. Since there are  $2^{-\xi}n - 2$ such clusters, we obtain
\begin{equation}
 \langle s \rangle_o = \frac{n}{2^{-\xi}n - 1} >  \frac{n^{1-\beta}}{2}.
\end{equation}
In the thermodynamic limit, the mean number of nodes per cluster diverges at the onset of the phase transition. This exact calculation thus shows that our criterion is also valid in a model in which only a finite number of cluster sizes are present. 
\newline\newline
In sum, under general conditions a percolation process  will be explosive if the mean number of nodes per cluster diverges at the onset of the phase transition in the thermodynamic limit. Equivalently, if the cluster density $n_c/n$ vanishes in the thermodynamic limit prior to the critical point, the transition is explosive. Although the criterion is deduced with a specific ansatz for the cluster-size distribution, simulations and analytical calculations on various explosive models indicate a more general validity. 
\begin{acknowledgments} We thank J.O. Indekeu for discussions and the anonymous referees for their useful suggestion. H.H. is supported by the Fonds voor Wetenschappelijk Onderzoek - Flanders (FWO-Vlaanderen). 
\end{acknowledgments}

\end{document}